# Implementations of ICT Innovations: A Comparative Analysis in terms of Challenges between Developed and Developing Countries

**Mutaz M. Al-Debei**
Assistant Professor
Department of Management Information Systems
The University of Jordan, Amman-Jordan
M.AlDebei@ju.edu.jo

**Enas M. Al-Lozi**
Assistant Professor
Department of Management Information Systems
Al-Zaytoonah University, Amman, Jordan
E.Al-Lozi@zuj.edu.jo

**ABSTRACT**

The main aim of this paper is to achieve a depth of understanding of the various similarities and differences in terms of challenges between developed and developing countries and in regards to the implementation of ICT innovations. Indeed, advances in Information and Communication Technologies (ICTs) have brought many innovations to the field of Information Systems (IS). Despite agreements on their importance to the success of organizations, the implementation processes of such innovations are multifaceted and require proper addressing of a wide-spread issues and challenges. In this study, we address this matter by first; synthesizing a comprehensive body of recent and classified literature concerning five ICT initiatives, second; analyzing and classifying ICTs challenges for both developed and developing countries as well as justifying their similarities and differences following thematic analysis qualitative methods, and third; presenting the study conclusions and identifying future research areas drawn upon the conducted comparative analysis.

## 1. Introduction

Recent advances in Information and Communication Technologies (ICTs) have brought many innovations to the field of Information Systems (IS). The main goal of such innovations, for both developed and developing countries, is the improvement of organizations' performance and the achievement of competitive advantages. Developing countries, in particular, are looking forward to achieve several social, economical and strategic gains by implementing various "western-originated" ICT initiatives. However, the overlapped implementation of these innovations spans over years, and is substantially associated with business logic strategic changes. Hence, they impose significant implications over organizations on the long run.

Many prescriptions to successfully employ these initiatives have been proposed in the literature. However, it seems that a great deal of customization effort is necessary to be done over these proposed prescriptions for the best fit within different organizations' internal/external environments. Variables such as 'organization size', 'sector', 'organizational and national cultures', 'politics', 'laws and regulations', and 'economic conditions' have been founded to be extremely influential on the implementation of these innovations.

Although implementation challenges are perceived to be to some extent universal, still there are unique features characterising each part of the world in regards to the implementation of ICT innovations and this is mainly due to the environmental differences. Indeed, ICT innovations not only deal with technologies and information content, but also their deliverables are shaped by the associated social/cultural context. This highlights the importance of identifying the implementation challenges of ICT innovations for developed and developing countries. Moreover, exploring the reasons behind the similarities and differences in terms of challenges between both types of nations is substantial to enhance our understanding of what constitute the most successful implementation of ICT innovations. To the best of our knowledge, this is the first paper tackling the differences in terms of challenges between developed and developing countries and in regards to different ICT innovations.

The rest of this study is organized as follows. Next, the methodology followed in this research for the purpose of achieving its goals and objectives is discussed. Thereafter, five ICT innovations are fully examined in section 3. For each initiative, normative and critical arguments and cases cornering both developed and developing countries have been synthesized and neutrally examined. Similarities and differences in terms of challenges have been pinpointed and vindicated in section 4. Finally, drawn conclusions and further research avenues have been proposed in section 5.

## 2. Research Design and Method

For the purpose of finding the similarities and differences between developed and developing countries in regards to the implementation of ICT innovations, a qualitative methodology is followed in this study. The primary source for data collection purposes in this study is defined as the related literature. A comprehensive body of recent and classified literature regarding the implementation of five ICT innovations (i.e. Enterprise Resource Planning "ERP", Business Process Re-engineering "BPR",

Inter-organizational Systems and E-Business, Information Systems Sourcing, and Knowledge Management) is systematically reviewed and synthesized.

Relevant articles reflecting the implementation of the identified ICT innovations in both developed and developing countries were extracted from electronic libraries and databases (e.g. ScienceDirect, EBSCO, JSTOR, and Scopus, and GoogleScholar), by means of keywords. The lists of references within the selected articles were also useful. In the process of selecting the body of literature to be covered in this study, we find it also important to cover normative, interpretive, and critical schools of thought so as to incorporate various perspectives in this regard and in turn to end up with neutral bias-free findings and conclusions.

Our literature search identified a sufficient and rich pool of papers addressing the implementation of the identified ICT innovations in both developed and developing countries. Indeed, about 120 references are cited and listed at the end of this study. This was not a surprise given the large volume of publications about Enterprise Resource Planning "ERP", Business Process Re-engineering "BPR", Inter-organizational Systems and E-Business, Information Systems Sourcing, and Knowledge Management in the field of information systems. Having the pool of literature created, one author (MMD) reviewed the titles and abstracts of the extracted papers for relevance to the current study. Papers were included in this study in accordance with the following dimensions collectively:

1- *Time Horizon*; that is covering the period from 1990's to 2010.

2- *Comprehensiveness*; that is covering normative, interpretive and critical schools of thought.

3- *Balance;* that is having fairly similar number of articles regarding each ICT initiative, and approximately similar number of articles in each ICT initiative addressing developed and developing countries.

4- *Quality of Articles;* in terms of content, number of citations, indexing, and publication source.

5- *Learning;* that is including articles that conducted comparisons in terms of implementation between developed and developing countries and in regards to the identified ICT innovations.

After selecting articles from literature to be included in this study, a thematic analysis of the content of these articles was started. Thematic analysis is a widely-used qualitative analytic method (see Roulston, 2001). Thematic analysis can be defined as a method for identifying, analysing, and reporting classified patterns (themes) within data (Braun and Clarke, 2006). Such an analysis organises and describes data sets in rich detail. However, it also can further interprets various aspects of the research topic (Boyatzis, 1998).

In this study, thematic analysis was applied separately on each ICT initiative. The authors of this study were both involved in this process and played the roles of thematic analyzers and evaluators. For example, if one author is the thematic analyzer for one ICT innovation, the other author is the evaluator and vice versa. However, during thematic analysis, all data sharing similar embedded meanings that indicated the same themes were coded and grouped together in a separate cluster. Each cluster is actually represents a shared similarity or difference between developed and developing countries and in regards to one the five ICT innovations included in this study. We have also made sure that generated clusters are mutually exclusive and unique. Then, generated clusters were provided with meaningful labels and appropriately justified. Finally a comparative analysis was conducted between these clusters so as to find similarities and differences in terms of challenges and in regards to each of the covered ICT innovations.

## 3. The Implementation Challenges of ICT Innovations in Developed and Developing Countries

As a first step in achieving the study goals and objectives, this section discusses the included five ICT initiatives (See Figure 1). For each initiative, the underlying organizational benefits that can be obtained due to the implementation of such an initiative are highlighted. Further, suggested approaches for successfully implementing each of the ICT innovations are synthesized. Moreover, issues according to both normative and interpretive/critical schools of thought are presented and supported by case studies extracted from relevant literature. Indeed, for comparison purposes in terms of challenges, case studies concerning both developed and developing nations are also provided.

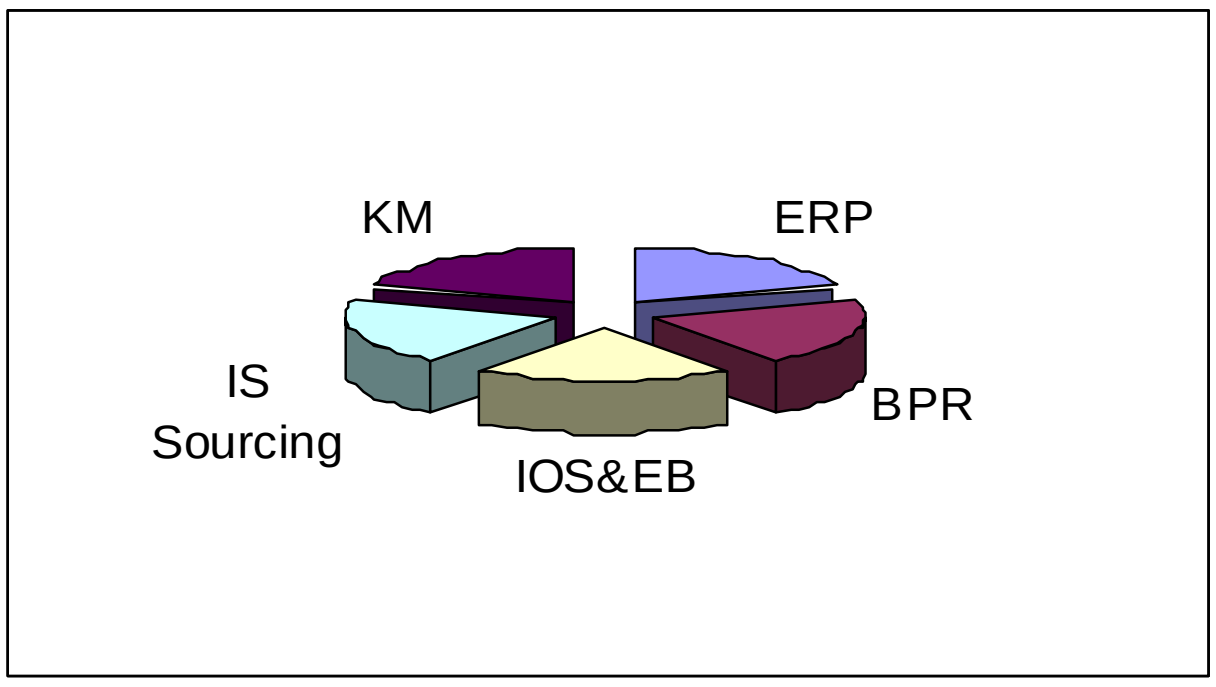

Figure 1. ICTs covered in the Study

### 3.1 Enterprise Resource Planning “ERP” Initiative

An ERP enterprise application is a standardized software package that integrates information across an organization. The implementation of an ERP software package in an organization represents organizational strategic continuum change (Bancroft, 1998), that transforms into a ‘radical’ one if Business Process Re-engineering (BPR) is considered. However, such an initiative fundamentally addresses operational integration within organizations. Through ERP implementations; organizations have sought to improve their competitiveness (Mabert et al., 2003), improve the quality of information and timeliness, reduce operating costs, replace legacy systems (Avison and Malaurent, 2007), enhance data visibility (Parr and Shanks, 2000), and improve management control over complex business processes (Davenport, 1998).

In this regard, the normative school of thought is concerned with finding the 'right model' of success for implementing ERP software packages. Ngai et al. (2008) and based on reviewing literature tackling the critical success factors (CSFs) in the implementation of enterprise resource planning (ERP) across 10 different countries/regions found that 'appropriate business and IT legacy systems', 'business plan/vision/goals/justification', 'business process reengineering', 'change management culture and programme', 'communication', 'ERP teamwork and composition', 'monitoring and evaluation of performance', 'project champion', 'project management', 'software/system development, testing and troubleshooting', 'top management support', 'data management', 'ERP strategy and implementation methodology', 'ERP vendor', 'organizational characteristics', 'fit between ERP and business/process', 'national culture' and 'country-related functional requirement' were the commonly extracted factors across these 10 countries/regions. They also found that in these 18 CSFs, 'top management support' and 'training and education' were the most frequently cited as the critical factors to the successful implementation of ERP systems.

On the basis of the applied analysis in this study, we found that awareness of the following factors constitutes the recipe for ERP implementation success: 'Sufficient financial resources', 'data accuracy', 'ERP strategy alignment with the competitive strategy', 'proper project management', 'effective change management', 'appropriate risk management', 'top-management support', 'effective structure and composition of teams', 'end-user training and involvement', 'powerful technological infrastructure', 'the selection and support of external consultants and vendors', and finally 'concentration on the decision made between BPR or ERP customization' (See Markus et al. ,2000 ; Sarker and Lee, 1999; Akkermans and Helden, 2002; Umble et al., 2003; Dezdar and Sulaiman, 2009; Metrejean, 2010).

However, it has been acknowledged in the related literature that organizations of different sizes approach ERP implementations differently (Markus et al., 2000; Luakkanen et al., 2007). For example, Buonanno et al. (2005) found that organizations from different markets face different issues and challenges when implementing ERP systems. In a similar manner, Parr and Shanks (2000) identified that requirements of ERP implementation success differ across ERP implementation approaches. Sheu et al. (2004) suggested that 'culture and language', 'management style', 'government/corporate policies', 'regulation/legal requirements', 'internal technical personnel resource/labor skills', and 'geography/time zone' represent national differences affecting ERP implementation practices across nations.

On the other hand, the interpretive/critical school of though argues that ERP implementation is a complex process since the use of technology is shaped by its social context. Thus applying the global template or recipe for implementation in different organizations operating in different environments without taking each local environment into account is very risky and may lead to catastrophic results. Accordingly, this school of thought argues that the process of ERP implementation needs to consider the following in order to be successful: 'Dynamics of organizational control', 'politics and power', 'the effect of cultural issues',

'different levels of experience and awareness', and 'the extent to which ICTs can support ERP implementation'. Based on a case study approach, Al-Mashari and Zairi (2000) identified the following factors for ERP implementation failure: 'Underestimating the human resources element in change', 'lack of cultural preparation', 'insufficient resources' (skills, manpower and finance), 'ineffective management of consultants', and 'separating IT from business affairs'. Xue et al. (2005) identified the following eight factors as influential ones leading to ERP failures in China: 'Language', 'report and table format', 'BPR', 'economic reform impact', 'cost-control system', 'human resource problem', 'price issues', and 'partnerships with ERP service organizations'.

As for the differences between developed and developing countries in regards to the implementation of ICT innovations, Huang and Palvia (2001) compared the implementation of ERP innovation amongst six developed and three developing countries. Interestingly, they found that when implementing ERP systems, developing countries face additional challenges related to *economical, political, cultural,* and *infrastructural* issues. More specifically, they indicated that low 'IT maturity', 'small firm size', 'lack of BRP experience and process management', 'long-term strategy', and 'poor project management experiences' were the main internal organizational challenges in developing countries. In another research and based on a comparison of ERP adoption among USA, UK and Greece, it has been found that Greek companies' 'internal culture', 'available resources', 'skills of employees', and 'management perception about ERP' have played crucial roles in determining ERP success or failure (Koh et al., 2006). Similarly, Avison and Malaurent (2007) identified that 'language and communication', 'governance concerning attitudes and values toward control and management', and 'laws and regulations' are cultural factors influencing the possible success of ERP systems.

Furthermore and by comparing two ERP cases, one in Australia and another in China, Shanks et al. (2000) found out that most ERP implementation CSFs differs because of the greater power-distance and collectivist nature of the Chinese culture. Tarafdar and Roy (2003) also argued that issues experienced by organizations in developing countries when implementing ERP systems are significantly different from those faced by organizations in the developed ones; mainly due to differences in 'the sophistication of IT use', 'culture', and 'social contexts'. Sheu et al. (2004) suggests that 'culture & language', 'management style', 'government/corporate policies', 'regulation/legal requirements', 'internal technical personnel resource/labour skills' and 'geography/time zone' represent national differences that affect ERP implementation practices across nations.

### 3.2 Business Process Re-Engineering (BPR) Initiative

BPR consists on a ' 'fundamental rethinking and radical redesign of business processes to achieve dramatic improvement in critical, contemporary measures of performance such as cost, quality, service and speed' ' (Hammer and Champy, 1993). In fact, BPR has been introduced by Hammer (1990), and Davenport and Short (1990) as an approach to increase operational efficiency and produce radical improvements in performance through streamlined business processes. According to Hammer and Champy (1993), ‘customer diversity and power’, ‘sever competition’, and ‘rapid environmental changes’ are the three reasons underlying the implementation of BPR. However, BPR initiative suggests that organizations need to elevate the importance of team-working and cross-functional processes to be appropriately integrated.

Many prescriptions have been proposed to manage this kind of radical changes at the organizational level. Hammer (1990) identified the following key principles to BPR: 'organize around outcomes, not tasks', 'have those who use the output of the process perform the process', 'subsume information-processing work into the real work that produces the information', 'treat geographically dispersed resources as though they were centralized', 'link parallel activities instead of integrating their results', 'put the decision point where the work is performed', 'build control into the process' and finally 'capture information once and at the source'. In the same vein, Davenport (1992) suggested that ‘developing the business vision and process objectives’, ‘identifying the business processes to be redesigned’, ‘understanding and measuring the existing processes’, ‘identifying IT levers’, and finally ‘designing, building and prototyping the new processes’ are the main factors constituting the right approach to manage BPR. For successful BPR implementations, Day (1994) identifies three principles to be practiced by senior management before carrying out a reengineering change as 'emphasizing external objectives', 'coordinating the activities & culture change' and 'making the information available to all team members'. Based on the experiences of consultants, Bashein et al. (1994) concluded the following CSFs of BPR: ‘Senior management commitment and sponsorship’, ‘realistic expectations’, ‘empowered and collaborative workers’, ‘strategic context of growth and expansion’, ‘shared vision’, ‘sound management practice’, ‘appropriate full-time participants’, and ‘sufficient budget’.

Interestingly, Zairi and Sinclair (1995) listed the following factors according to their importance to BPR success in a descending order (from the most important to the least): 'leadership', 'team make-up', 'available IT expertise', 'project targets', 'customer focus', 'existing IT systems', 'project’s time-frame', 'process knowledge', 'change management', 'communication', 'management systems', 'performance measurement', 'training', 'organization structure', 'organization culture' and 'investment'. Hall et al. (1993) proposed the following as three critical determinants of BPR success: 'breadth' (concerning the number of business units involved), 'depth' (the change to six organizational elements; organizational structure, roles & responsibilities, measurement & incentives, IT, shared values and skills), and 'leadership' (concerns the extent of top-management support). Maull et al. (2003) identified that 'taking a strategic approach', 'integrating performance measurement', 'creating business process architecture', 'involving human & organizational factors' and finally 'identifying the role of IT' constitute a smooth road to BPR success.

Critically, Davenport and Stoddard (1994) identified seven myths of BPR as follows: ‘the myth of reengineering novelty’, ‘the myth of clean state’, ‘the myth of IS leadership’, ‘the myth of reengineering vs. quality’, ‘the myth of top-down design’, ‘the myth of reengineering vs. transformation’, and ‘the myth of reengineering’s permanence’. Reviewed literature reveals that BPR has been widely misunderstood as it has been equated to downsizing and/or client-server computing (Malhotra, 1998). Mumford (1996) warned from personal risks since BPR is associated with downsizing. Moreover, Knights and McGabe (1998) identified that BPR may bring stress to employee nature of work, thus increase their resistance and lead to disruption of organizational goals and objectives.

Betroni et al. (2009) argued that BPR projects very often fail to meet the inherently high expectations of re-engineering. Indeed, surveys and studies estimate the percentage of BPR failures to be as high as 70% (Hammer and Champy, 1993). Malhotra (1998) indicated that 'lack of sustained management commitment and leadership', 'unrealistic scope and expectations', and 'resistance to change' are the principal obstacles that business process engineering initiatives face. On the other hand, Sarker and Lee (1999) identified 'lack of detailed knowledge about functional areas', 'hidden agendas of top-management', 'lack of knowledge of (and over-reliance on) computer-based BPR tools', 'poor choice of metaphors in organizational language', and 'lack of communication' as the main reasons behind BPR failure at US TELECO. Based on three case studies, Willcocks and Smith (1995) suggest that BPR is too often method-driven, and argued that it needs further holistic attention where human, social, and political issues are considered as BPR main inhibitors.

As for developing countries, Ranganathan and Dhaliwal (2001) identified that 'lack of human and financial resources', 'lack of internal IT expertise and capabilities', and 'lack of champion for BPR initiatives' as the main problems faced by the Singapore organizations. Salman (2004) identifies the following impediments faced by organizations in developing countries when carrying out BRP initiatives: 'Lack of holistic view', 'lack of sponsorship', 'unsound financial condition', 'sense of complacency', 'failure to distinguish between BPR and other improvement programs', 'burying the BPR project among corporate agenda', 'resistance to change', 'communication gap', 'generic low usage of technology and sceptical attitude towards technology', and 'lack of experience in handling BPR projects'. Furthermore, Abdolvand et al. (2008) and based on reporting survey results indicating BPR readiness in two Iranian companies found that 'resistance to change' is one of the main factors leading to failures in regards to the implementation of BPR initiative. On the other hand, Abdolvand et al. (2008) study found that egalitarian leadership, collaborative working environment, top management commitment, supportive management, and use of information technology are factors leading to successful implementations of BPR initiatives in Iran. In a similar vein, Nasierowski (2000) indicated that BPR necessitate to be repackaged to fit the particulars of the Mexican cultural, social, and financial settings. Nasierowski also pointed out that success heavily depends on the strong support of the highest political levels and business commitment. Saxena (1996), Based on providing a number of reengineering examples carried out in Indian public sector, claimed that for a successful BPR in developing countries, a holistic treatment and attention should be set to strategy, organizational structure, IT, and culture.

### 3.3 Inter-organizational Systems and E-Business Initiative

The fundamental message of IOS&EB innovation is that through its implementation an organization would improve its profit, achieve strategic sustainability (Chircu et al, 2000), cut costs, run business smoothly and effectively (Lu, 2003), build customer base quickly (Kauffman and Walden, 2001), reshape customer and supplier relationships, and streamline its business processes (Daniel and Grimshaw, 2002). Turning into dotcoms, organizations have been promised to be able to create value to their customers characterized by low-cost and ease-of-use (see Timmers, 1999; Amit and Zott, 2001). Further, Porter (2001) argued that a flood of new entrants has come into many industries since the Internet has reduced barriers to entry. For example, based on interviewing Michael Dell, Magretta (1998) described how Dell Company benefits from it virtual integration through exploiting ICT to blur the traditional boundaries in the value chain (disintermediation). Moreover, Kraemer and Dedrick (2002) depicted how Cisco Company established its IOS to implement its strategic focus and to leverage its virtual organization, leading to higher growth and profit rates.

Similarly, Ghosh (1998) argued that “Companies that do not want to participate in Internet commerce may be forced to do so by competitors or customers”. This seems to be true and planning approaches to implement this kind of ICT innovations have been proposed by many scholars. For example, Leitch and Warren (2003) suggested the following stages for successful dot.com implementations: 'strategic & business evaluation', 'system analysis & design', 'systems e-commerce design', 'implementation' and finally 'post-implementation'. Choucri et al. (2003) developed a model that represents the CSFs for dotcoms based on ‘Access’ (infrastructure and services), ‘Capacity’ (social factors, economic factors, and policy factors), and ‘Opportunity Penetration’. Jackson and Sloane (2007) argued that for organizations to operate successfully, ‘integration among organizational processes’, ‘human resources’, ‘organizational culture’, and ‘sound management’ is required.

Furthermore, Cullen and Taylor (2009) study yielded five composite factors that are perceived by users to influence successful e-commerce use. The proposed critical success factors for e-business initiative, according to their study are: “System quality,” “information quality,” “management and use,” “world Wide Web – assurance and empathy,” and “trust”. Of these and according to the study of Cullen and Taylor (2009), all respondents ranked information quality, system quality, and trust as being of most importance, but differences in the rankings between purchasing and selling respondents are evident. Another study by Sharkey et al. (2010), and on the basis of Delone and Mclean Model, found significant relationships between Information Quality and System Quality and three success dimensions of e-commerce initiatives: intention to use, user satisfaction and intention to transact. The study also found the following information and system quality constructs to be most important in predicting e-commerce success: ease of understanding, personalisation and reliability. In particular, they found that reliability is more important than usability where transactions are concerned and security is important to transactional zones of e-commerce systems. on the other hand, it has been also found that factors such as ‘internet access and its infrastructure’, ‘user confidence’, ‘trust’, ‘security’, and ‘privacy’ are claimed to downsize e-business growth (Connolly, 1998).

Interestingly, it has been proven that organizations from different size and/or operating in different industry sectors may adopt different e-commerce strategies as they usually face different issues and challenges in regards to the implementation process (*See* Doherty et al., 2001; Daniel et al., 2003). It has been also proven the value proposition characteristics do matter (Brynjolfsson et al., 2000) as implementation challenges do vary to some extent according to the nature of offerings of different organizations. It is also worthwhile mentioning that the electronic market witnessed a number of high-profile dot.com collapses (Howcroft, 2001). Indeed, Shapiro (2000), for example, warned from liberty erosion as one of the negative e-commerce implications. Shapiro raises issues concerning control and privacy and their social implications. Moreover, Howcroft (2001) discussed the following myths concerning the dot.com market: ‘the myth of the new economy’, ‘the myth of success’, ‘the myth of the entrepreneurial geek’, ‘the myth of the level playing field’, ‘the myth of innovation’, ‘the myth of the virtual’, and ‘the myth of the online shopping experience’.

Using the case of Egypt, Ghoneim et al. (2001) concluded that an institutional role is needed to regulate e-commerce. Markus et al. (2001) concluded that organizational use of Internet in some parts of Asia differs from that in the USA in the following aspects: 'financial infrastructure', 'legal and regulatory infrastructure', 'national policies, telecommunication infrastructure', 'language and education', and finally 'organizations size, structure and their control systems'. A case based research conducted by Kshetri (2007) classified barriers to e-commerce in developing countries, into 'economic', 'socio-political', and 'cognitive' at both consumer and business levels. Based on a survey in Sri Lanka, Kapurubandara and Lawson (2007) identified that implementations of e-commerce technologies is inhibited by both 'internal factors' (owner/manager characteristics, firm characteristics, and cost and return on investment) and 'external ones' (infrastructure, social and cultural, political, legal and regulatory). Based on examining i-metal Chinese case study, Hempel and Kwong (2001) indicated that EB implementations in developing countries presents unique challenges that are not presented in developed ones. Not only have they identified infrastructure challenges in terms of financial, legal, and physical deficiencies, but also they have identified different business philosophies and culture to be the most significant challenge that influences company-marketplace relationship.

Empirically through surveying 95 Jordanian companies from different sectors, Al-Debei and Shannak (2005) found out that the main reasons underlying the very small number of e-commerce applications by Jordanian companies are: 'national-cultural' (lack of awareness about e-commerce and its requirement), 'security & privacy issues', and 'lack of regularity & legislative bodies'. In a similar vein, Salman (2004) indicated that 'lack of basic automation in place', 'poor management skills', 'lack of e-commerce integration', 'economical situations', 'politics', 'inadequate infrastructure', 'pressing digital divide', 'feeble human capital', 'culture' and 'society' represent inhibitors to implement e-commerce technologies in developing countries. In Thailand, the results of Sebora et al. (2009) study confirmed that the achievement orientation and locus of control of founders and business emphasis on reliability and ease of use functions of e-service quality are positively related to the success of e-commerce entrepreneurial ventures in Thailand.

### 3.4 IS-Sourcing Initiative

The handing-over of tasks within the IS functions to a third-party is known as IS Sourcing. This kind of innovation is intended to offer organizations tremendous performance enhancements such as significant cost reductions (Nicholson and Sahay, 2001), effective use of human resources and exploitation of external advanced technologies, ability to focus on core competencies, reduction in capital investments, and challenging the rational planning view. However, Loh and Venkatraman (1992) noted that organizations' tendency to copy successful practices of others without taking into consideration their own attributes are one of the major reasons behind IS-sourcing besides the incapability to manage internal IS resources which represents another key reason for failure.

In fact, IS sourcing can be classified differently based on the dimension taken into account. In fact, IS-sourcing could be handled either internally; known as insourcing, or externally known as outsourcing (See DiRomualdo and Gurbaxani, 1998). Moreover, it could be classified as total vs. selective (See Heinzl, 1993), single vendor vs. multiple vendors (See Currie, 1998), joint venture vs. strategic partnership (See Fitzgerald and Willcocks, 1994), and on-shoring vs. off-shoring (See Shao and David, 2007). Interestingly, these types are not mutually exclusive as there are overlapping aspects amongst them. Another classification of IS-Sourcing was provided by Millar (1994) who classified IS-Sourcing in four categories as follows: general, transitional, business process, and business benefit contracting.

Catalogues of key factors influencing IS sourcing successful implantations have been advocated. Heeks et al. (2001), for example, revealed the following challenges concerning IS-Sourcing: 'distance', 'essential tacit knowledge & informal information transfer', and finally 'different cultural values'. On the other hand, Aron et al. (2005) suggested a taxonomy for IS-sourcing risks as 'strategic risks' (behaviour of both parties), 'operational risks' (i.e. communication), 'intrinsic risks of atrophy' (losing core people and expertise), and 'intrinsic risks of location' (i.e. geopolitical, socio-political, and cultural). Further, Issues such as transfer of assets and information security have been discussed by Lee (1996) as inhibitors of IS-Sourcing. As for developed countries in particular, Shao and David (2007) argued that IT workers in developed countries are negatively influenced by the IS-sourcing trend. They predicted that those able to interlink their IT skills with business needs are those having the best chance to survive. Mclaughlin (2003) assured the same point; Mclaughlin indicated that IS-sourcing not only has changed the US landscape for software professionals, but also has reshaped companies' project planning and the criteria based on which they select their employees. Moreover and based on a German telecommunication Company's experience that setting up a satellite operation in India, Kobitzsch et al. (2001) identified that 'legal', 'knowledge transfer', 'development and project management', 'quality management', 'language', 'time', and 'infrastructure' as critical factors in the arena of IS-Sourcing.

Outsourcing in particular seems to be very challenging and risky undertaking, despite its promising benefits. Willcocks et al. (1995), for example, suggested a formula to mitigate risks associated with outsourcing consisting of the following principles: 'market logic not management despair', 'rationalization not rationing', 'commodities not differentiators', and finally 'targeted not total'. Using Rational Exchange Theory, Goles and Chin (2005) identified 'commitment', 'consensus', 'cultural compatibility', 'flexibility', 'interdependence', and 'trust' as attributes of outsourcing success. Further, they identified processes of success in this regard as 'communication', 'conflict resolution', 'coordination', 'cooperation' and 'integration'. In a study that addresses e-commerce and ASP outsourcing, Lee et al. (2003) recommended the following guidelines for achieving an informal effective partnership when outsourcing is considered: 'understand each other business', 'prioritize short-term and long-term goals', 'define realistic expectations', 'share benefits and risks', 'develop performance standards', 'expect changes and revisions', 'prepare for the unexpected', and 'nurture the relationship'. However, the applied analysis in this study reveals that challenges of IS-sourcing not only differ across its type and scope, but they also differ across various organizational strategies, sizes, and sectors (See Sobol and Apte, 1995; Gallaugher and Stoller, 2004). On the other hand, Qu et al. (2010) interestingly found that IT insourcing is more effective for developing IT-enabled business processes (IEBP), which subsequently lead to superior firm performance.

Based on a study that included comprehensive literature review as well as conducting interviews with Offshore Software Development (OSD) experts, a list of CSFs was developed from the perspective of German-speaking companies by Remus and Wiener (2009). They have actually identified the following seven CSFs as generally being the most relevant for the successful implementation of an OSD project: 'definition of clear project goals'; 'continuous controlling of project results'; 'ensuring of a continuous communication flow'; 'high quality of offshore employees'; 'good language abilities of the offshore employees in German and English'; 'composition of an appropriate project team'; and 'preparation of a detailed project specification'. From an on-going longitudinal study of British organizations offshore to India, Nicholson and Sahay (2001) concluded that 'culture asymmetries', 'organizational politics' (resources of power), and 'time/space dispersion' represent substantial challenges to British organizations' management. Using transaction theory, Qu and Brocklehurst (2003) found that Chinese legal system forms a main difficulty in reaching a competitive situation in offshore outsourcing. On the other hand, Khan et al. (2002) indicated that the role of Indian government plays major roles in enhancing its infrastructure and that the role of virtual around the clock approaches is substantial to outsourcing initiative success.

On the basis of delineating the success of one Vietnamese company as an outsourcing partner, Gallaugher and Stoller (2004) indicated that Vietnamese potential growth is limited by the subsequent issues: 'government policies', 'technology infrastructure', and 'comparative limitations in terms of population and size'. Using the quantitative-qualitative approach, Khalfan and Alshawaf (2003) concluded that both 'environmental' (educational status, national IT strategy, economic status, political system, technological status, and legal status) and 'cultural' challenges (language/communication, religion, and behavioural norms and attitudes) face Kuwaiti public sector when outsourcing is addressed. Coward (2003) identified that 'culture', 'language', and 'time zone differences' are critical issues faced by American SMEs when tackling outsourcing. Coward also explained that political stability of the provider's country represents an important factor to American SMEs in their outsourcing decision.

### 3.5 Knowledge Management Initiative

Providing the means for managing knowledge as one of organizations main possessions in order to leverage its intellectual capital is the primary significance underlying KM initiatives. By leveraging KM, organizations have hunted to improve their competitiveness over time (Nonaka and Takeuchi, 1995) as KM intends to signify a structure that facilitates problem-solving, decision-making, and strategic planning. In fact, knowledge management encompasses identifying, capturing, selecting, organizing, disseminating, and transferring knowledge across boundaries. However, making knowledge visible, manageable and transferable is a multifarious process due to its tacit nature (See Nanoka, 1994), its stickiness (See Von Hippel and Tyre, 1996), and its nature of being distributed (See Tsoukas, 1996). Accordingly, Carlile (2002) argued that knowledge is both a source of and a barrier to innovation since it is localized, embedded, and invested in practice.

Through surveying more than 300 senior executives, Gold et al. (2001) found that organizations would achieve effective KM not only by focusing on knowledge process architecture (acquisition, conversion, application, and protection), but also through focusing on the knowledge infrastructure (technology, structure, and culture). Moreover, Ajmal et al. (2010), based on their study that examined the critical success factors for KM initiatives in project-based organisations, identified six CSFs in this context: ‘familiarity with KM’; ‘coordination among employees and departments’; ‘incentive for knowledge efforts’; ‘authority to perform knowledge activities’; ‘system for handling knowledge’; and ‘cultural support’. Empirically, Alvai (1999) indicated that even though 'technology' represents one of the key concerns regarding KM; 'cultural', 'managerial', and 'informational issues' represent the substantial ones. Critically and based on examining two case studies; Ebank and Brightco, Swan et al. (1999) compared cognitive network model; where technology is perceived as the CSF, with community networking model; where trust and collaboration is perceived as the CSF. Accordingly, they warned from the potential impact of the former one on KM within organizations. In a similar manner, Walsham (2001) indicated that ICTs are beneficial only if carefully used to support the development and communication of human meaning. In a similar vein, Reid and Slazinski (2003) indicated that 'cultural heterogeneity' of participants in a service learning program at Purdue University represents issues to their KM initiative. Moreover and through a series of semi-structured interviews, Desouza and Evaristo (2003) indicated that 'change management' and 'culture' are relevant issues pertaining to KM across borders.

As for developing countries, in particular, Nguyen and Johanson (2008), through field work interviews, found out that the main reason hampers Vietnamese from accepting elements of a knowledge nation is their society long-standing features (economic, social, cultural, and political). Using a case of Indian pharmaceutical organizations, Kale and Little (2005) indicated that building-up new competencies in developing countries is complex undertaking due to political and economic convolutions. Using the case of Sudan, Ghobrial (2006) indicated the following problems confronting Sudanese society when KM partnership: 'reduced access to knowledge', 'lack of regional-global integration and cooperation', 'lack of ICT cultural-ethics', 'lack of human-resources', 'lack of education', 'lack of ICT development structure program', 'lack of financial resources', 'society poverty and illiteracy', 'legislation gaps', 'lack of ICT policy guidelines', and 'insecurity and bureaucracy aligned with ICT'. Based on examining the development of a knowledge portal at the Housing and Development Board (HBD) in Singapore, Teo (2005) identified the HBD encountered challenges related to 'people' and 'culture', 'process', 'content management', and 'technology'. Generally speaking, Malhan and Gulati (2003) indicated that developing countries confront 'technological', 'socio-economical', 'attitudinal', 'geographical', and 'linguistics' challenges in having access to knowledge. They also indicate that despite the presence of sparkling ideas and facts in India, lack of interests in knowledge activities expressed by top-management wastes these innovative ideas away.

## 4. Discussion: Comparative Analysis

The results of the applied analysis in this study shows that the implementation of ICT innovations is a complex course of action; mainly because of the essentially needed consistent interactions amongst people, technology, and business processes. In fact, this unique mixture makes the success of implementation of various ICT innovations way far from being straightforward and somehow unpredictable. The applied analysis also reveals that organizations, in general, face many substantial challenges throughout the implementation of such innovations, and that challenges along with their extent vary across different organizations and in accordance with their macro and micro features and characteristics. Hence, we argue that there is no one recipe of success in regards to the implementation of ICT innovations that fits all organizations operating in various environments.

Indeed, as the relevancy of ICT innovations to developing countries has been assured (Walsham et al., 2007), the debate regarding the challenges of ICT innovations has become much more substantial. From the previous section, it has become clear that there is an increasing research interest in delineating the differences between developed and developing countries in terms of implementation challenges of ICT innovations. Our analysis in this study assures that both developed and developing countries share common issues and challenges in the micro-organizational level in regards to the implementation of ICT innovations, as summarized in Table 1.

| Issues and Challenges |
|---|
| Lack of clear boundaries among different ICTs. |
| The soft nature of ICT projects and its consequences. |
| The rationality of ICT (s) implementation decision and other related decisions. |
| Aligning the ICT innovation strategy with the competitive strategy of an organization. |
| Lack of concern of the human-resource element (needed skills, involvement and retention). |
| Lack of top-management commitment and support. |
| Ineffective management of consultants. |
| Resistance to change. |
| The dynamics of power (organizational politics). |

Table 1. Common ICT Implementation Challenges for both Developed and Developing Countries

Our analysis reveals that ICT innovations are not mutually exclusive as significant overlapping areas are available, in addition to the high level of interdependence amongst most of them (See Figure 2). In fact, each ICT initiative undergoes from the disappearance of clear perimeters and its overlapping aspects during implementation with other ICT innovations. To give just a general overview, BPR is highly considered as a pre-requisite for ERP which is in turn considered a backbone for IOS&EB innovations. KM is considered significant in managing organizational knowledge, while other initiatives and their related issues constitute a significant portion of that knowledge. On the other hand, IS-sourcing is considered a choice for organizations confronting a decision regarding the implementation of all other ICT innovations. Hence, we believe that ICT innovations generally complement each other rather than acting as substitutes, despite the fact that they all nearly hold the same message towards organizations entitled as “enhancing the performance and providing a competitive advantage”.

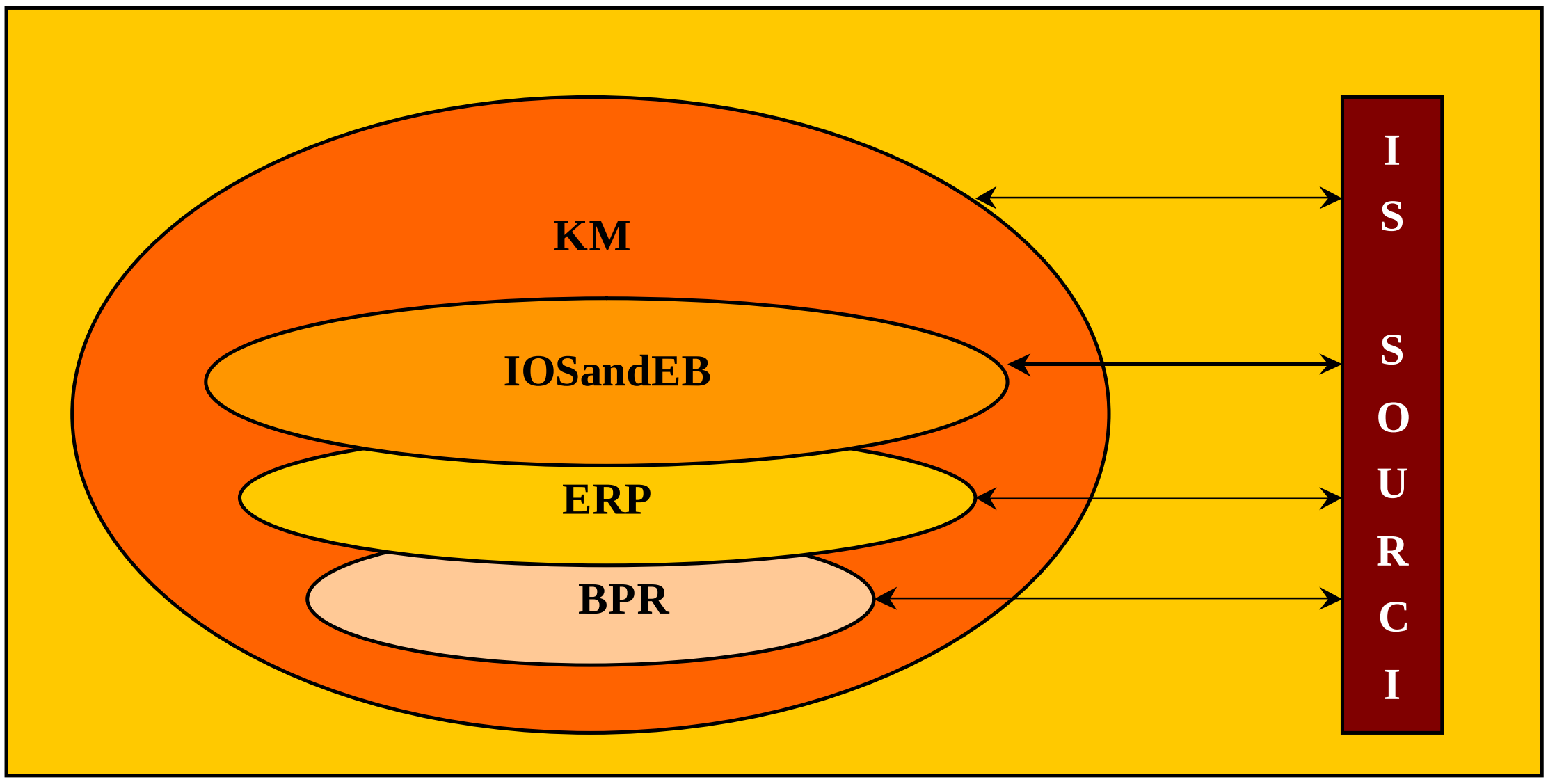

Figure 2. Interdependences amongst ICT Innovations

Another common challenge is actually related to the nature of ICT projects given that such projects are normally less tangible than others and thus their implementations are significantly more challenging than other projects in terms of requirement specifications, control, monitoring, and evaluation. Making decisions in regards to the implementation represents another common challenge (See Table 2 and Table 3). Indeed, there is a need to choose innovations helpful in achieving organizational goals and objectives so as to be implemented. Consequently, organizations are required to decide whether these innovations are to be implemented internally or should be outsourced. Moreover, decisions regarding the approach to be followed for implementation purposes and functional areas to be covered should be made. In addition to these general decisions, there are specific decisions to be made and in accordance with the chosen innovation for implementation (See Table 3). For example, if ERP innovation is to be implemented, an organization is required to make decisions in regards to whether Big-Bang, Middle-Road, or Vanilla approach should be followed. On the other hand, if KM innovation is to be implemented, there is a need to decide whether personalization or codification is the appropriate strategy to be followed in a specific organization.

| Decisions | Issues and challenges |
|---|---|
| When and which ICT innovation(s) to implement? | **A.** The significance of the ICT innovation(s). |
| | **B.** Aligning the ICT innovation strategy with the organization competitive strategy. |
| Should we implement the selected ICT innovation(s) in-house or should we source it? | The decision whether to consider IS-sourcing or not. |
| Which vendor(s) and/or consultant(s) should we select? | The selection of the vendor(s)/consultant(s) needed for the implementation of the selected ICT innovation(s) in order to acquire human and technology recourses. |
| Which implementation approach or strategy should we adopt and follow? | The selection of the implementation approach in regards to the selected ICT innovation(s). |
| What is the scope of the selected ICT innovation(s)? | The selection of the functional areas or business units to be covered and supported by the selected ICT innovation(s). |

Table 2. Common Decisional Challenges for both Developed and Developing Countries

In fact, many critical decisions in this regard are normally made based on informal communications managers do with their social network. In some other times, managers make decisions based on their tendency to copy successful firms in terms of ICTs implementations and without studying the feasibility and need of such innovations to their organizations. Such practices would lead organizations to make irrational decisions that in turn may lead to catastrophic results; given that organizations, in this case, neglect the appropriateness of the chosen innovation(s) to their settings. Hence, we believe that ICT innovations should be implemented at the right time and should fit an organization's strategy, structure, management-style, and most importantly the culture, if that innovation is to be implemented successfully.

Table 3 below lists other decisional issues and challenges related to each ICT innovation covered in this study. These challenges are shared between developed and developing countries.

| ICT Innovations | Issues and Challenges |
|---|---|
| **BPR** | The number of functional areas or business units involved. |
| **ERP** | The selection amongst Big-Bang Middle-Road, or Vanilla as an implementation approach. |
| | The selection between BPR or ERP software customization for the purpose of ERP fit with the business processes of an organization. |
| **IOS&EB** | The degree of separation/integration |
| **IS-sourcing** | The selection between insourcing vs. outsourcing. |
| | The selection between total vs. selective sourcing. |
| | The selection between single vendor vs. multiple vendors. |
| | The selection among IS-sourcing joint venture, and strategic partnership. |
| | The selection between on-shoring vs. off-shoring. |
| **KM** | The selection between personalization or codification strategy for knowledge transfer. |

Table 3. ICT-Specific Challenges for both Developed and Developing Countries

For both developed and developing countries, there are also other important challenges for the implementation of ICT innovations. Both nations also suffer from lack of concern of human resource element. Often in such a context, organizations neglect or underestimate the role of human resources along with their skills and capabilities as organizations normally perceive such projects as purely ICT ones. Believing that acquiring the best technology assures its deployment success is one of the immense pitfalls. The implementations of such ICT innovations are usually considered as strategic changes that affect organizations' on the long run since they stroke every single aspect of the entire organization. Hence, underestimating the human-resource elements during strategic changes associated with the implementations of ICT innovations plays a significant role in hindering their implementations success. For successful implementations, we argue that organizations should assure having all needed resources for such initiatives. This includes financial, human-resources, technological, technical, business skills and other relevant resources. Factors such as users' involvement, training, retentions, and powerful team formation as well as strong leadership facilitate successful implementations of ICT innovations. Lack of top-management support decreases the perceived value of ICT innovations; thus decreasing the level of commitment of various users, team members, and other stakeholders.

As clarified earlier, the implementation of ICT innovations is complex undertaking. Hence, organizations often acquire the needed skills and expertise to successfully implement such innovations by incorporating consultants in this process. However, the cost associated with consultants is usually high, and the value they can add to the process of the implementation is hardly assured. This is because, most often, the organization and the acquired consultants share different interests and goals. Consultants sometimes tend to complete the project as fast as possible and move on to the next one, whilst quality and effective completion of the project is main concern for organizations. Indeed, the applied analysis in this study reveals that ineffective management of consultants is one of the main challenges organizations confront when implementing ICT innovations. This challenge is actually owed to poor communication and knowledge transfer problems between consultants and relevant members from organizations, in addition to the tendency of consultants to maintain their power, influence, and organization-dependency.

| Categories | Issues and Challenges |
|---|---|
| **Infrastructural** (Physical, Legal, HR, Technological, and others) | Lack of appropriate technological infrastructure |
| | Lack of local vendors & consultants for ICT innovations |
| | Lack of internet accessibility |
| | Expensive internet and other technological services |
| | Lack of R&D resources and facilities |
| **Economical** | Unfavorable economic status and Recession |
| | Low economic growth |
| | Low income level |
| | Scarcity of opportunities |
| **Political** | Unfavorable laws, regulations and policies |
| | Lack of governance |
| | Lack of intellectual property laws |
| | Unstable environment |
| **Cultural** | Poor education system in some countries |
| | Lack of national awareness & perception concerning |
| | ICTs. |
| | Low IT maturity |
| | High resistance to change |
| | Low English proficiency |
| | Low IT diffusion |
| | Low level of innovation |
| **Geographical** | Geographical distance |
| | Different time zones |

Table 4. Macro-Environmental (National) Challenges of ICT Implementations in Developing Countries

Another critical challenge facing both developed and developing organization in regards to the implementation of ICT innovation is 'resistance to change'. This normally happens when, for example, some members of the organizations become reluctant to acquire new skills and techniques associated with the new innovation. It some other times, resistance to change happens when some influential members of the organization stand offensively against the implementation of the new ICT innovation. In fact, the underlying reasons behind resistance to change are numerous. For example, some employees would resist the change associated with the implementation of the new ICT innovation because of the fear of the unknown; accordingly they prefer to keep the status qua which they have a fair control over and got used to. Others would resist the change associated with the new ICT innovation because they believe this would lead them to lose their power, accumulated experience, control over processes, or jobs in the organization. In some other cases, members of an organization would show resistance because they believe that the new innovation would strengthen the overseeing and monitoring capabilities of their managers and profoundly change the dynamics of power within the organization.

In addition to the common challenges discussed above, there are extra unique ones facing only organizations operating in developing countries. This can be traced back to the fact that ICT innovations were firstly established for developed countries as they were originated in the western part of the world. Therefore, the degree of compatibility of these ICT innovations with variables related to macro and micro environments of an organization operating in developed countries is substantially higher than those operating in developing countries. Indeed, there are substantial differences mainly in political, economical, social/cultural, and infrastructural (i.e. technological, physical, legal, human resources) aspects between developed and developing nations (See Table 4). Nonetheless, listing these factors exclusively for developing nations does not indicate their ultimate absence in the developed ones, but we sort them exclusively in this study because these challenges are more generalizable and evident within developing countries.

| Categories | Issues and Challenges |
|---|---|
| **Infrastructural** | Inadequate technological infrastructure at the |
| (Physical, Legal, HR, Technological, and others) | organizational level |
| | Lack of financial recourses and capabilities |
| | Lack of needed skills and expertise within organizations |
| **Economical** | Hierarchal management style |
| | Power-distance |
| | Strict line-of-orders |
| | Lack of R&D resources and facilities within organizations |
| | Relatively small-size of organizations |
| **Political** | The dynamics of power |
| | Arbitrary and social-driven Decision-making |
| | Hidden management agendas |
| | The fear of losing jobs |
| **Cultural** | Inappropriate technological culture |
| | low IT maturity |
| | Lack of IT skills and experience |
| | Unfavorable attitudes and values toward control and change |
| | Inadequate English proficiency |
| | Interlinked social and organizational relationships |

Table 5. Micro-Environmental (Organizational) Challenges of ICT Implementations in Developing Countries

Organizations within developing countries not only experience challenges at the macro-national level, but they also face issues and challenges related to the micro-organizational environment (See Table 5). Indeed, some issues have been inherited from the macro-national level such as low IT maturity and diffusion, inadequate technological infrastructure which is the backbone for such ICT innovations, weak financial situations due to the recession, lack of R&D resources and facilities, lack of relevant expertise, lack of awareness about the added-value associated with ICT innovations along with their significance, poor English proficiency that is needed for effective communications with consultants and vendors, and geographical location issues. Such additional challenges definitely obstruct successful implementations of various ICT innovations in developing countries.

Other challenges within developing countries could be traced back to (a) the hierarchal style that exhibits a strong "command and control" management style; thus hinders innovation, (b) interlinks between organizational and social relationships (For instance, in developing countries, the social status of an individual usually affects his/her status in the organization. Moreover, people normally mix between their social and professional relations), and (c) lack of proper planning and modelling of the business.

Organizational politics represents also a critical challenge facing developing countries in regards to the implementation of ICT innovations. Indeed, ICT innovations usually change the way business are conducted, which consequently affects the dynamics of power within organizational settings and lead to a higher level of resistance. To give just one example, employees who have expertise in using an existing legacy system will be threatened of losing their knowledge power if the organization decides to replace the legacy system with a new ICT innovation.

In addition to all challenges discussed earlier, there are also other related challenges in this context which are mainly associated with the nature of ICT innovations. In fact, the ICT innovations which are covered and examined in this study are regarded as enterprise "large-scale" Innovations. Being classified as enterprise innovations indicates their high total cost of ownership and the kind of expertise they require. Hence and given the size and the financial conditions of most organizations operating in developing countries in addition to the modest availability of essential skills and expertise, the ability to successfully implement such demanding ICT innovations significantly declines.

## 5. CONCLUSIONS

The underlying reason behind the implementation of ICT innovations is gaining competitive advantages; primarily through enhancing the performance, effectiveness, and efficiency of organizations. Nonetheless, the implementation of ICT innovations is complex undertaking and usually requires a powerful alignment amongst technology, business, and human-related factors. The applied analysis in this study reveals that both developed and developing countries undergo several challenges in regards to the implementation of ICT innovations. Lack of concern about the human-resources, lack of top-management support, ineffective management of consultants, resistance to change, and the dynamics of power within an organization are just few examples of those challenges. Interestingly, the analysis also reveals that developing countries face extra unique challenges in this context and in both macro and micro levels such as infrastructure inadequacy, economics and politics instability, culture unsuitability, time/location dispersion, and management style inappropriateness.

Moreover, the analysis shows that the relevancy and successful implementation methods of ICT innovations significantly vary across nations and organizations. Retrospectively, we advocate that ‘one size fits all’ is not an applicable approach in the arena of ICT innovations and that there is no one universal recipe for success that can be applied everywhere, but a rationale scientific approach is recommended to be followed by managers when confronting decisions regarding the implementation of ICT innovations given their significant influences on the on-going value of organizations.

Broadly speaking, the analysis also found that there is no direct proportional relationship between the implementation of ICT innovations and organizational outcomes. This is because such a relationship is mediated by the social context factor which is mainly derived from the cultural settings. Thus, we believe that successful implementations of ICT innovations are those creating a balance between conflicting requirements and are performed in an adaptive and customized manner with various micro-organizational and macro-national variables.